\documentclass[runningheads]{llncs}
\usepackage[T1]{fontenc}
\usepackage{graphicx}

\usepackage{tabularx}
\usepackage{booktabs}
\usepackage{xcolor}

\begin{document}
%



\title{Enhancing DR Classification with Swin Transformer and Shifted Window Attention}
%
%
\author{Meher Boulaabi\inst{1}\orcidID{0009-0000-6773-2781} \and
Takwa Ben Aïcha Gader\inst{3}\orcidID{0000-0002-3786-3649)} \and 
Afef Kacem Echi\inst{3}\orcidID{0000-0001-9219-5228)}
Zied Bouraoui \inst{2}\orcidID{0000-0002-1662-4163}}
\authorrunning{M. Boulaabi et al.}
%
\institute{Unversity of Tunis, ENSIT-LaTICE Lab - University of Monastir, FSM 
\and
Artois University, CRIL-CNRS, France
\and 
University of Tunis - ENSIT-LaTICE Lab., Tunisia
}
\maketitle              

\begin{abstract}
Diabetic retinopathy (DR) is a leading cause of blindness worldwide, underscoring the importance of early detection for effective treatment. However, automated DR classification remains challenging due to variations in image quality, class imbalance, and pixel-level similarities that hinder model training. To address these issues, we propose a robust preprocessing pipeline incorporating image cropping, Contrast-Limited Adaptive Histogram Equalization (CLAHE), and targeted data augmentation to improve model generalization and resilience.
Our approach leverages the Swin Transformer, which utilizes hierarchical token processing and shifted window attention to efficiently capture fine-grained features while maintaining linear computational complexity. We validate our method on the Aptos and IDRiD datasets for multi-class DR classification, achieving accuracy rates of 89.65\% and 97.40\%, respectively. These results demonstrate the effectiveness of our model,  particularly in detecting early-stage DR, highlighting its potential for improving automated retinal screening in clinical settings.

\end{abstract}
\section{Introduction}
Diabetic Retinopathy (DR) is a microvascular complication of diabetes and the leading preventable cause of blindness among the working-age population worldwide. Early detection and timely intervention are essential to preventing vision loss.
The emergence of deep learning revolutionized DR detection by enabling automated feature learning directly from raw images. 
Transformer-based architectures have recently emerged as powerful tools in medical image analysis, offering a compelling alternative to traditional deep learning models. Vision Transformers (ViTs), in particular, leverage self-attention mechanisms to capture long-range dependencies, enabling rich representations of both local and global features. However, standard ViTs are limited by their quadratic computational complexity with respect to image size, making them inefficient for high-resolution medical images.
To overcome this limitation, we employ Swin Transformers, which introduce a hierarchical architecture with shifted window-based self-attention. This approach reduces complexity to linear with image size, enhancing computational efficiency for large-scale visual data. The shifted windows also improve feature extraction by effectively modeling both local details and global context, while maintaining scalability.
For diabetic retinopathy (DR) classification, we implement a comprehensive pipeline that includes dataset preparation, preprocessing, and augmentation. We apply image cropping to eliminate irrelevant background regions, reducing noise and enhancing focus on clinically relevant features. Contrast-Limited Adaptive Histogram Equalization (CLAHE) is used to improve local contrast and highlight subtle structures, aiding feature extraction. To address class imbalance, we apply data augmentation techniques such as rotation and horizontal flipping, enriching the training data and improving generalization. Together, these steps contribute to more accurate and robust DR classification across varying severity levels.

\section{Methodology}
Our model's architecture is illustrated in Figure \ref{model}. 
The Swin Transformer serves as the backbone. Unlike ViTs, which process images in a global self-attention framework, the Swin Transformer applies shifted window-based self-attention, reducing computational complexity and improving efficiency for high-resolution medical images. Below, we outline its key components:

\begin{figure}[t]    
\small
    \centering
    \includegraphics[width=0.7\linewidth]{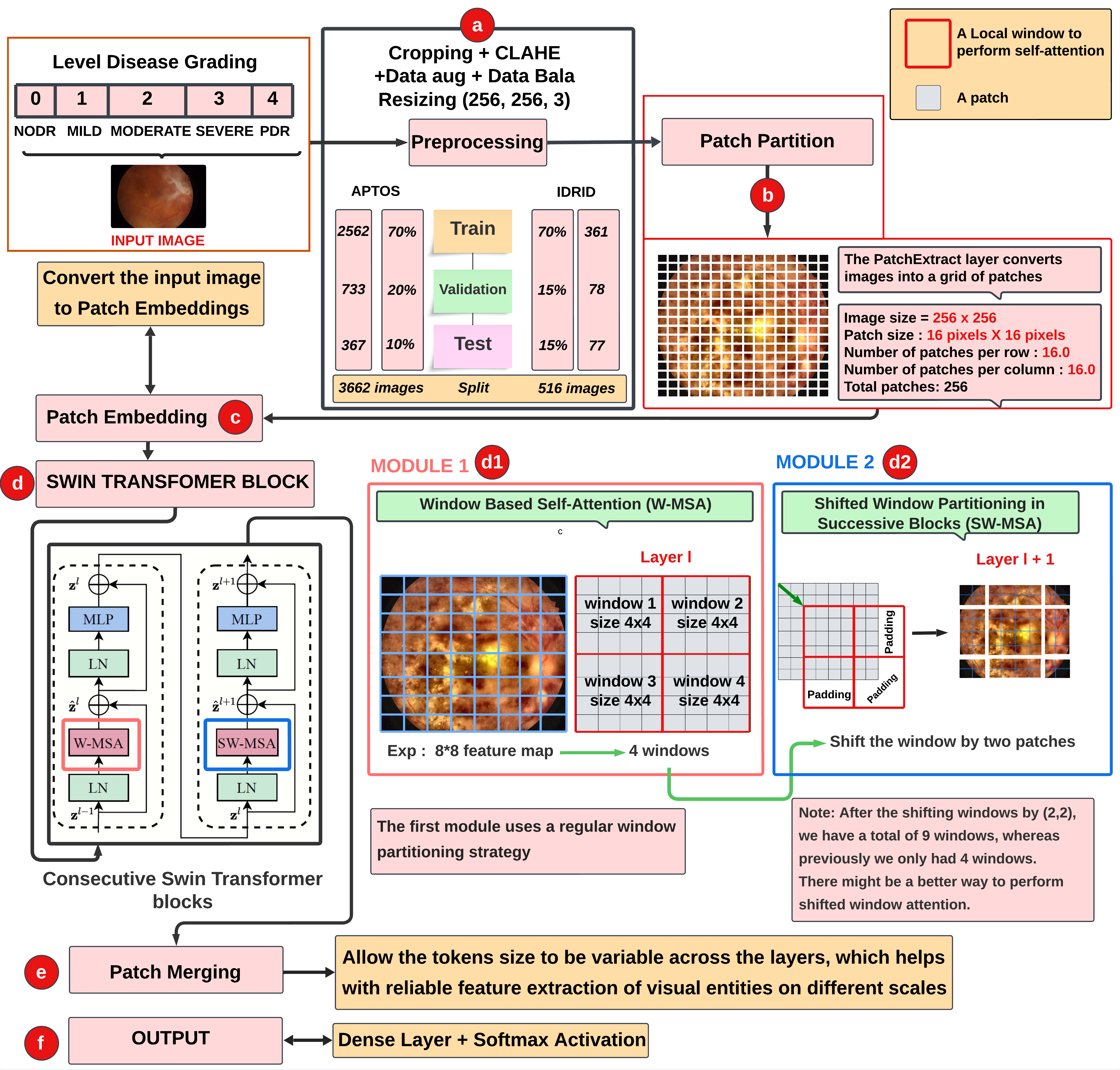}
    \caption{DR classification pipeline. The input images undergo preprocessing with CLAHE and data augmentation to enhance robustness. Images are divided into patches using PatchExtract and projected into a high-dimensional space via PatchEmbedding. These embeddings pass through transformer blocks, where Module 1 applies standard window-based self-attention and Module 2 uses shifted window attention for better cross-window learning. Residual connections and MLP layers with dropout ensure stability and improved feature extraction. Finally, PatchMerging reduces spatial dimensions, followed by global average pooling and a dense layer with softmax classifier.}
    \label{model}
\end{figure}


\paragraph{Swin Transformer Block:} Each consists of a shifted window-based Multi-Head Self-Attention (MSA) module, followed by a two-layer Multi-Layer Perceptron (MLP) with a GELU activation function. Each MSA module and MLP is preceded by a LayerNorm (LN) layer, and residual connections are incorporated to improve gradient flow and prevent vanishing gradients.
\paragraph{Window-Based Self-Attention (W-MSA):} 
Unlike traditional self-attention mechanisms, W-MSA operates within local non-overlapping windows, significantly reducing computational overhead.  Given an image of size \( h \times w \) with each window containing \( M \times M \) patches, the computational complexity for global MSA versus window-based MSA is
    $\Omega(\text{MSA}) = 4hwC^2 + 2(hw)^2C$ and 
    $\Omega(\text{W-MSA}) = 4hwC^2 + 2M^2hwC$    
    where \(C\) is the embedding dimension. Global self-attention computation, as in the standard Vision ViT, is computationally expensive for large \(hw\), whereas window-based self-attention is more scalable.
    
\paragraph{Shifted Window Partitioning in Successive Blocks (SW-MSA).} 
In consecutive blocks, the first module applies regular window partitioning, while the second module shifts the windows, allowing for improved feature continuity and better spatial representation. The forward propagation for two consecutive Swin Transformer blocks is given as
    $\hat{z}^l = \text{W-MSA}(\text{LN}(z^{l-1})) + z^{l-1}$,    $z^l = \text{MLP}(\text{LN}(\hat{z}^l)) + \hat{z}^l$,
    $\hat{z}^{l+1} = \text{SW-MSA}(\text{LN}(z^l)) + z^l$, and 
    $z^{l+1} = \text{MLP}(\text{LN}(\hat{z}^{l+1})) + \hat{z}^{l+1}$,
where \( z^{l-1} \) denotes the input feature representations from the preceding layer.  A challenge with shifted window partitioning is the increased number of windows in the second configuration, some of which are smaller. This affects spatial dependency capture. To mitigate this, \textit{cyclic shifting} pads edge windows with neighboring regions, ensuring feature continuity and improving spatial processing. This shifted partitioning strategy improves the ability of the model to capture long-range dependencies while maintaining computational efficiency.

\section{Experiments}

This section analyzes the performance of our model against baseline methods on the APTOS and IDRiD datasets, widely used for diabetic retinopathy detection.

\paragraph{Datasets.}
We work on grading DR using the IDRiD and APTOS datasets, focusing on classifying the five severity levels. These include level 0 (no apparent retinopathy), level 1 (mild non-proliferative diabetic retinopathy), level 2 (moderate), level 3 (severe), and level 4 (proliferative diabetic retinopathy, or PDR). Although the two datasets differ in terms of image quality, quantity and characteristics, they share the same grading scale. We applied the same classification algorithm to both datasets and introduced variations to ensure robust performance and generalizability across different data sources.





\paragraph{Data Preprocessing.}

To enhance image quality and improve model performance, we applied a consistent preprocessing pipeline to the IDRiD and APTOS datasets 
First, image cropping was performed to remove extraneous background elements while preserving the essential circular structure of the eye, ensuring the model focused on relevant retinal features. Next, CLAHE was used to enhance contrast and visibility, particularly in areas with irregular illumination, improving feature clarity in DR images. Both datasets were then organized into training, validation, and testing sets.
To improve generalization, data augmentation techniques, including random rotations (up to 360 degrees) and horizontal flips, were applied to simulate real-world imaging variability and reduce overfitting risks 
Table \ref{preprocessing_comparison} highlights the impact of preprocessing, demonstrating its role in enhancing classification by addressing pixel similarity and image quality variability. 


\begin{table}[t]
\scriptsize
\centering
\caption{Performance on the IDRiD dataset with and without preprocessing.}
\label{preprocessing_comparison}
\begin{tabular}{lccccccc}
\toprule
\multirow{2}{*}{Class} & \multicolumn{3}{c|}{W/o preprocessing} & \multicolumn{3}{c|}{W/ preprocessing} \\ \cline{2-7} 
                       & Precision & Recall & F1-Score            & Precision  & Recall & F1-Score            \\ \midrule
Grade 0                & 0.45      & 0.38   & 0.41                  & 0.96       & 1.00   & 0.98                \\ \midrule
Grade 1                & 0.00      & 0.00   & 0.00                  & 1.00       & 1.00   & 1.00                \\ \midrule
Grade 2                & 0.37      & 0.78   & 0.51                  & 1.00       & 0.92   & 0.96                \\\midrule
Grade 3                & 0.67      & 0.21   & 0.32                  & 0.93       & 1.00   & 0.97                \\ \midrule
Grade 4                & 1.00      & 0.08   & 0.14                  & 1.00       & 1.00   & 1.00                \\ \bottomrule
\end{tabular}
\end{table}

\paragraph{Training.} The model is trained using a patch size of (16,16) and a learning rate of \(1 \times 10^{-3}\) and a batch size of 32 for 200 epochs. Softmax activation and Cross-Entropy loss are used for classification, with the AdamW optimizer and early stopping set to 15 epochs.




\paragraph{Results.}
Table \ref{metrics} highlights the model’s strong specificity and sensitivity, confirming its suitability for real-world clinical applications. It also demonstrates the effectiveness of our model and preprocessing pipeline in accurately classifying all DR severity levels, from grade 0 to grade 4. Additionally, Table \ref{comp_tab} shows that preprocessing plays a critical role in boosting performance, with notable improvements over models trained without it. These results further underscore the potential of Swin Transformers for medical image analysis, showcasing their precision and reliability in clinical settings.

\begin{table}
\scriptsize
\centering
\caption{Class-wise Precision, Recall, and F1-Score for Aptos and IDRiD Datasets}
\label{metrics}
\begin{tabular}{lccc|ccc}
\toprule
\textbf{Class} & \multicolumn{3}{c|}{\textbf{Aptos}} & \multicolumn{3}{c}{\textbf{IDRiD}} \\ \midrule
 & \textbf{Precision} & \textbf{Recall} & \textbf{F1-Score} & \textbf{Precision} & \textbf{Recall} & \textbf{F1-Score} \\ \midrule
Grade 0 & 0.94 & 0.98 & 0.96 & 0.96 & 1.00 & 0.98 \\ \midrule
Grade 1 & 0.86 & 0.68 & 0.76 & 1.00 & 1.00 & 1.00 \\ \midrule
Grade 2 & 0.82 & 0.89 & 0.86 & 0.92 & 0.92 & 0.92 \\\midrule
Grade 3 & 0.81 & 0.68 & 0.74 & 0.92 & 0.86 & 0.89 \\\midrule
Grade 4 & 0.88 & 0.77 & 0.82 & 1.00 & 1.00 & 1.00 \\\midrule
Overall & 0.86 & 0.80 & 0.83 & 0.96 & 0.96 & 0.96 \\ \bottomrule
\end{tabular}
\end{table}

\begin{table}[t]
\centering
\scriptsize
\caption{Comparison of models for multi-class classification.}
\label{comp_tab}
\begin{tabular}{lcc}
\toprule
 \textbf{Method} & \textbf{Dataset} & \textbf{Accuracy (\%)} \\ \midrule
 Blended features + DNN \cite{qummar2019deep} & Aptos & 80.96 \\  \midrule
Deep dual-branch model (transfer learning) \cite{shakibania2024dual}&  Aptos & 89.60 \\  \midrule
\textbf{Our Method} & \textbf{Aptos} & \textbf{89.65} \\  \midrule
 SqueezeNet classifier \cite{uppamma2023diabetic} & Idrid & 94.20  \\  \midrule
 CNN, SqueezeNet, ResNet-50, Inception-V3, DFTSA-Net, CLAHE \cite{nazir2020diabetic} & Idrid & 95.00  \\ \midrule
 \textbf{Our Method} & \textbf{Idrid} & \textbf{97.40} \\  \bottomrule
\end{tabular}
\end{table}

\section{Conclusion}
We introduced a Swin Transformer-based model for DR classification, addressing class imbalance, pixel similarity, and subtle retinal variations. By integrating a comprehensive preprocessing pipeline with image cropping, CLAHE, and augmentation, we enhanced feature extraction and model performance. 

\section*{Acknowledgments}
This work was supported by ANR-22-CE23-0002 ERIANA.

%

\bibliographystyle{splncs04}
\bibliography{ref}

\end{document}